%% file: sorters.tex
\documentclass[conference]{IEEEtran}
\usepackage{cite}

\ifCLASSINFOpdf
  \usepackage[pdftex]{graphicx}
\else
\fi
\usepackage[cmex10]{amsmath}
\usepackage{amsthm}
\usepackage{amssymb}

\usepackage[vlined,linesnumbered,ruled]{algorithm2e}
\SetAlFnt{\small}

\usepackage{threeparttable}

\usepackage{url}

\usepackage{enumerate}

\usepackage{tikz}
\usetikzlibrary{arrows,decorations.pathreplacing}

\input{notation}

\input{definitions}

\begin{document}
%
\title{On Metric Sorting for Successive Cancellation List Decoding of Polar
Codes}

\author{%
  \IEEEauthorblockN{%
    Alexios~Balatsoukas-Stimming\IEEEauthorrefmark{1}, %
    Mani~Bastani~Parizi\IEEEauthorrefmark{2}, and %
    Andreas~Burg\IEEEauthorrefmark{3}%
  }%
  \IEEEauthorblockA{%
    EPFL, Lausanne, Switzerland\\ 
    Email: \IEEEauthorrefmark{1}alexios.balatsoukas@epfl.ch,
    \IEEEauthorrefmark{2}mani.bastaniparizi@epfl.ch,
    \IEEEauthorrefmark{3}andreas.burg@epfl.ch%
  }%
}


%


\maketitle
\begin{abstract} 
  We focus on the metric sorter unit of successive cancellation list decoders
  for polar codes, which lies on the critical path in all current hardware
  implementations of the decoder. We review existing metric sorter architectures
  and we propose two new architectures that exploit the structure of the path
  metrics in a log-likelihood ratio based formulation of successive
  cancellation list decoding. Our synthesis results show that, for the list
  size of $L=32$, our first proposed sorter is $14\%$ faster and $45\%$ smaller
  than existing sorters, while for smaller list sizes, our second sorter has a
  higher delay in return for up to $36\%$ reduction in the area.
\end{abstract}

%
\IEEEpeerreviewmaketitle

\section{Introduction}
Polar codes \cite{Arik09} are a recently introduced class of provably capacity
achieving channel codes with efficient encoding \emph{and decoding} algorithms.
Successive cancellation list (SCL) decoding \cite{Tal11} is a decoding algorithm
which improves upon the conventional successive cancellation decoding of polar
codes in terms of frame error rate, while only increasing the decoding
complexity linearly by a factor of $L$, where $L$ is the \emph{list size} of the
decoder. The key step of SCL decoding is to choose $L$ paths with the smallest
path metric out of $2L$ possible paths. While theoretically this problem can be
solved by finding the median of the $2L$ metrics whose computational complexity
is $O(L)$ \cite[Section 9.3]{cormen2001introduction}, in practice, since $L$ is
relatively small, it is easier to just sort the $2L$ numbers and pick the first
$L$.

The \emph{metric sorter} (which implements the above-mentioned sorting step)
turns out to be the crucial component of all the SCL decoder hardware
architectures proposed in the literature~\cite{Bala14,icassp,Lin14,Yuan14,Zhang14b}. In
\cite{Bala14,icassp}, the metric sorter lies on the critical path of the decoder
for $L \geq 4$, so it determines the maximum clock frequency of the decoder. In
\cite{Lin14}, the sorter is pipelined in order to remove it from the critical 
path, but the number of cycles required to decode each codeword is increased.

\paragraph*{Contribution}
In this work, we review and compare the metric sorter architectures used in the
existing SCL decoder architectures in \cite{Bala14,icassp,Lin14}.
We then leverage the properties of the LLR-based path metric introduced in
\cite{icassp} in order to introduce two new sorter architectures; a \emph{pruned
bitonic} sorter and a \emph{bubble} sorter. By comparing the synthesis results
of the new sorters and the existing sorters we highlight an advantage of the
LLR-based formulation of the SCL decoder which leads to an optimized
implementation of the metric sorter.

\section{Problem Statement} 
Let $\bm = [m_0,m_1,\cdots,m_{2L-1}]$ denote the $2L$ real path-metrics to be
sorted. In an LL-based implementation of SCL decoding (e.g.,
in~\cite{Bala14,Lin14}), $\bm$ contains arbitrary real numbers and a
general sorting problem of size $2L$ needs to be solved. In an LLR-based
implementation \cite{icassp,Bala14tsp}, however, $\bm$ has a particular
structure that can be exploited to simplify the sorting task. 
More specifically, let $\mu_0 \le \mu_1 \le \cdots \le \mu_{L-1}$ be the $L$
sorted path metrics from the previous step of SCL decoding. Then, in an
LLR-based implementation, the $2L$ new path metrics in $\bm$ are computed as
\begin{equation*}
  m_{2 \ell }  := \mu_\ell \quad \text{and} \quad
  m_{2 \ell + 1}  := \mu_\ell + a_\ell, 
  \qquad \ell = 0,1,\dots,L-1,
\end{equation*}
where $a_\ell \ge 0$, for all $\ell \in \{0,1,\dots,L-1\}$. Thus, the
sorting problem is to find a sorted list of $L$ smallest elements of $\bm$ when
the elements of $\bm$ have the following two properties
\begin{subequations}%
  \begin{align}%
    m_{2 \ell} & \le m_{2 (\ell + 1)}, 
    \label{eq:oldListSorted} \\
    m_{2 \ell} & \le m_{2 \ell + 1}. 
    \label{eq:positiveIncrements}%
  \end{align}%
\end{subequations}%
for $\ell \in \{0,1,\dots,L-2\}$.

Note that \eqref{eq:oldListSorted} and \eqref{eq:positiveIncrements} imply that
out of $\binom{2L}{2} = L (2L-1)$ unknown pairwise relations between the
elements of $\bm$, $L^2$ are known (every even-indexed element is smaller than
all its following elements). Hence, we expect the sorting complexity to be
reduced by a factor of $2$. 
\begin{remark} 
  For SCL decoding, in order for the assumptions on the list structure to hold,
  besides the above mentioned problem, a general sorting problem of size $L$
  needs to be solved infrequently \cite{Bala14tsp}.  We note that this problem
  can be solved by using a sorter that finds the $L$ smallest elements of a list
  with properties \eqref{eq:oldListSorted} and \eqref{eq:positiveIncrements},
  $L-1$ times in a row.%
  \footnote{%
    Let $a_0, a_1,\dots, a_{L-1}$ be arbitrary real numbers. For $\ell =
    0,1,\dots,L-2$, set $m_{2\ell} := -\infty$ and $m_{2\ell+1} = a_\ell$.
    Finally set $m_{2L-2} := a_{L-1}$ and $m_{2L-1} := +\infty$. It is easy to
    check that \eqref{eq:oldListSorted} and \eqref{eq:positiveIncrements} hold
    for the list $\bm$ and the ordered $L$ smallest elements of this list are
    $[-\infty, -\infty, \dots, \min_{0\le\ell\le L-1} a_\ell]$. Thus, we can
    find the minimum of up to $L$ arbitrary real numbers using such a sorter.
    Consequently, using the sorter $L-1$ times in a row we can sort an arbitrary
    set of $L$  real numbers.%
  }
\end{remark}

\section{Existing Metric Sorter Architectures}\label{sec:existing}
\subsection{Radix-$2L$ and Pruned Radix-$2L$ Sorters}\label{sec:existingr2l}
In our previous work of \cite{Bala14,icassp} we used a radix-$2L$ sorter
\cite{Amaru12}, which compares every pair of elements $(m_\ell, m_{\ell'})$ and
then combines the results to find the $L$ smallest elements. This solution
requires $L(2L-1)$ comparators together with $L$ $2L$-to-$1$ multiplexers. 
In an LLR-based implementation of SCL decoding, the full radix-$2L$ sorter can
be \emph{pruned} by removing the comparators corresponding to $L^2$ known
relations and observing that $m_{2L-1}$ is never among the $L$ smallest
elements. Eliminating these unnecessary comparators, the number of required
comparators is reduced to $(L-1)^2$ and we only require $(L-1)$ multiplexers of size $(2L-2)$-to-$1$~\cite{Bala14tsp}.

\subsection{Bitonic Sorter}
The authors of \cite{Lin14} used a bitonic sorter~\cite{Batcher68}.  A bitonic
sorter that can sort $2L$ numbers consists of $(\log L + 1)$
\emph{super-stages}.  Each super-stage $s \in \{1,2,\dots,\log L + 1\}$ contains
$s$ \emph{stages}.  The total number of stages  is, hence,
\begin{equation} 
  s_{\mathrm{tot}}^{\mathrm{BT}} = \sum_{s=1}^{\log L +1} s = \frac{1}{2}(\log L
  + 1)(\log L + 2). \label{eq:stagesbt}
\end{equation}
The length of the critical path of the sorter is determined by the number of
stages.  Each stage contains $L$ compare-and-select (CAS) units consisting of
one comparator and a $2$-to-$2$ MUX. Thus, the total number of CAS units in a
bitonic sorter is 
\begin{equation} 
  c_{\mathrm{tot}}^{\mathrm{BT}} = \frac{L}{2}(\log L + 1)(\log L + 2).
  \label{eq:compsbt}
\end{equation}
An example of a bitonic sorting network that can sort $2L=8$ numbers is given in
Fig.~\ref{fig:bitonic}.


\section{Proposed Sorters}

\subsection{Pruned Bitonic Sorter}\label{sec:pbs}
\begin{figure}[tb]
  \centering
  \includegraphics[width=0.42\textwidth]{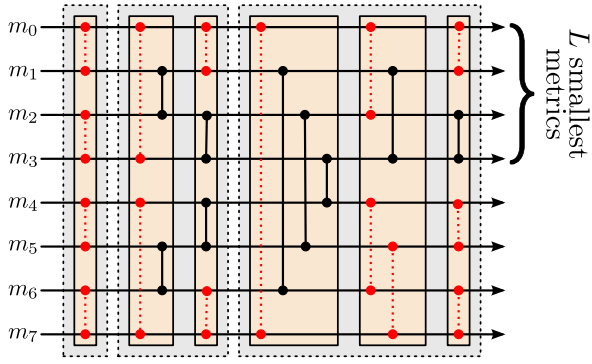}
  \caption{Bitonic sorter for $L=4$. Each vertical line represents a CAS
    unit that has the two endpoints of the line as inputs. The full bitonic
    sorter requires all the depicted CAS units, while in the pruned bitonic
  sorter all CAS units in red dotted lines can be removed.}
  \label{fig:bitonic}
\end{figure}

As was the case with the radix-$2L$ sorter, the known relations between 
the elements can be exploited to simplify the bitonic sorter. In particular, due
to \eqref{eq:positiveIncrements}, the results of all sorters in stage $1$ are
already known. Thus, stage $1$ can be removed completely from
the sorting network. Moreover, the result of all comparators whose one input is
$m_0$ are also known, since $m_0$ is, by construction, always the smallest
element of $\bm$. Furthermore, since $m_{2L-1}$ is never amongst the $L$
smallest elements of the list, all comparisons involving $m_{2L-1}$ are 
irrelevant and the corresponding CAS units can be removed. Finally, we can 
remove the $L/2$ last CAS units of the $\log L$ final stages of super-stage 
$\log L + 1$, since they are responsible for sorting the last $L$ elements 
of $\bm$ while we are only interested in its first $L$ elements. 
The unnecessary CAS units for $2L = 8$ are illustrated with dotted red 
lines in Fig.~\ref{fig:bitonic}.
%

Since only stage 1 is  completely removed from the sorting network, the
number of stages in the pruned bitonic sorter is, 
\begin{equation} 
  s_{\mathrm{tot}}^{\mathrm{PBT}} = s_{\mathrm{tot}}^{\mathrm{BT}}-1 =
  \frac{1}{2}(\log L + 1)(\log L + 2) - 1. 
\end{equation}
Therefore, the delay of the pruned bitonic sorter is only slightly smaller
than that of the full bitonic sorter, especially for large list sizes $L$.

To compute the number of CAS units in a pruned bitonic sorter, we note that the
first super-stage is eliminated completely. 
In all remaining super-stages except the last
one, the $2$ CAS units per stage that are connected to $m_0$ and $m_{2L-1}$ are
removed.
In the last super-stage, we can remove the CAS
units connected to $m_0$ plus all the CAS units in the second half of the last
$\log L$ stages since they contribute in sorting the $L$ largest elements of the
list.
Hence, the total number of CAS units in the pruned bitonic sorter can
be shown to be equal to 
\begin{equation} 
  c_{\mathrm{tot}}^{\mathrm{PBT}} = \Big(\frac L2 - 1\Big) (\log L) (\log L +2)
  + 1.  
\end{equation}
By examining the ratio between $c_{\mathrm{tot}}^{\mathrm{BT}}$ and
$c_{\mathrm{tot}}^{\mathrm{PBT}}$ we can conclude that, similarly to the maximum
delay, the relative reduction in the number of comparators also diminishes with
increasing list size $L$.

\subsection{Bubble Sorter}\label{sec:bubble_sorting}
While bubble sort \cite[Chapter~2]{cormen2001introduction} is in general an
inefficient sorting algorithm, it turns out to be a suitable candidate for our
particular problem. More precisely, properties
\eqref{eq:oldListSorted} and \eqref{eq:positiveIncrements} result in a specific
data dependency structure of the algorithm enabling an efficient hardware
implementation of the sorter. Furthermore, since we only require the sorted list
of $L$ smallest elements of $\bm$ (rather than sorting the entire list $\bm$) we
can simplify the sorter by only implementing the first half of the rounds of the
algorithm.

\subsubsection{Data Dependency}
The bubble~sort algorithm is formalized in Alg.~\ref{alg:bubbleSort}. It is
clear that Alg.~\ref{alg:bubbleSort} sorts the full list. By restricting
the \textbf{while} condition as ``exists $\ell \in \{0,1,\dots,L-1\}$ such
that $m_{\ell} > m_{\ell+1}$'' one can simplify the algorithm to only output the
first $L$ ordered elements of the list $\bm$.
\begin{algorithm}[htb]
  \While{exists $\ell$ such that $m_\ell > m_{\ell+1}$}{%
    \For{$\ell = 2L-1$ \KwTo $1$}{
      \If{$m_\ell < m_{\ell-1}$}{%
	Swap $m_\ell$ and $m_{\ell-1}$\; \nllabel{lin:swap}
      }
    }
  }
  \Return{$\bm$}
  \caption{The Bubble Sort Algorithm}
  \label{alg:bubbleSort}
\end{algorithm}
\begin{lemma} \label{lem:bubbleSortProperties} 
  Let $m_\ell^t$ denote the element at position $\ell$ of the list at \emph{the
  beginning of round $t$} of the \textbf{while} loop in
  Alg.~\ref{alg:bubbleSort} and, 
  \begin{equation} 
    \calB_t \triangleq \left\{\ell \in \{1,2,\dots,2L-1\}: m_\ell^t <
      m_{\ell-1}^t\right\}. \label{eq:btDef} 
  \end{equation} 
  Then \eqref{eq:oldListSorted} and \eqref{eq:positiveIncrements} imply that for
  all $t \ge 1$,
  \begin{enumerate}[(i)]
    \item $\calB_t$ does not contain adjacent indices,
      \label{itm:nonAdjacentB}
    \item the \textbf{if} body (line~\ref{lin:swap}) is executed at round $t$
      iff $\ell \in \calB_t$,
      \label{itm:forExec}
    \item $\calB_{t+1} \subseteq \calB_t + 1$, where $\calA + a \triangleq
      \{x+a: x \in \calA\}$.
      \label{itm:bPlusOne}
  \end{enumerate}
\end{lemma}
\begin{IEEEproof}
  To prove the lemma, we will prove that for all $t \ge 1$,
  \begin{equation}
    m_{\ell}^t \ge m_{\ell-2}^t \qquad \text{for all $\ell \in \calB_t$}.
     \label{eq:bIndexOrder}
  \end{equation}

  We first show that \eqref{eq:bIndexOrder} implies
  \eqref{itm:nonAdjacentB}--\eqref{itm:bPlusOne} and then prove
  \eqref{eq:bIndexOrder}.

  \begin{enumerate}[(i)]
    \item Suppose $\ell \in \calB_t$, hence, $m_\ell^t < m_{\ell-1}^t$ and
      \eqref{eq:bIndexOrder} implies $m_{\ell}^t \ge m_{\ell-2}^t$. Thus
      $m_{\ell-1}^t > m_{\ell-2}^t$ which implies $\ell-1 \not\in \calB_t$.
    \item  Note that the element at position $\ell$ of the list is changed iff
      line~\ref{lin:swap} is executed for indices $\ell$ or $\ell+1$.  
      
      We use strong induction on $\ell$ to prove \eqref{itm:forExec}. Clearly
      line~\ref{lin:swap} is executed for the first time for an index $\ell^* =
      \max \calB_t$.
  
      Assume line~\ref{lin:swap} is executed for some index $\ell$. This
      implies $m_{\ell} < m_{\ell-1}$ before execution of this line when
      $m_{\ell-1} = m_{\ell-1}^t$ (since line~\ref{lin:swap} has not been
      executed for $\ell$ nor for $\ell-1$ so far).  Now if $\ell+1 \not\in
      \calB_t$, then $m_\ell^t = m_\ell$ as well by the induction assumption
      (since line~\ref{lin:swap} is not executed for index $\ell+1$) hence
      $m_{\ell}^t < m_{\ell-1}^t$ which means $\ell \in \calB_t$. Otherwise,
      $\ell+1 \in \calB_t$, implies line~\ref{lin:swap} is executed for
      $\ell+1$.  Since  $\calB_t$ does not contain adjacent elements,
      line~\ref{lin:swap} is \emph{not} executed for $\ell+2$. This implies
      $m_\ell = m_{\ell+1}^t \ge m_{\ell-1} = m_{\ell-1}^t$ by
      \eqref{eq:bIndexOrder} which contradicts the assumption of loop being
      executed for $\ell$. 

      Conversely, assume $\ell \in \calB_t$. Since $\ell+1 \not\in \calB_t$,
      line~\ref{lin:swap} is not executed for $\ell+1$ by assumption. Hence once
      the \textbf{for} loop is executed for index $\ell$, $m_{\ell} = m_\ell^t$
      and (as we justified before) $m_{\ell-1} = m_{\ell-1}^t$. Therefore,
      $m_{\ell} < m_{\ell-1}$ and line~\ref{lin:swap} is executed for $\ell$.  
    \item Using \eqref{itm:nonAdjacentB} and \eqref{itm:forExec}, we can
      explicitly write the time-evolution of the list as:
      \begin{equation}
	m_{\ell}^{t+1} = \begin{cases}
	  m_{\ell-1}^t, & \text{if $\ell \in \calB_t$}, \\
	  m_{\ell}^t, & \text{if $\ell \not\in \calB_t$ and $\ell+1 \not\in
	  \calB_t$}, \\
	  m_{\ell+1}^t & \text{if $\ell+1 \in \calB_t$}. \\
	\end{cases}
	\label{eq:mGeneral}
      \end{equation}
      Therefore, if $\ell \in \calB_t$, $m_{\ell}^{t+1} = m_{\ell-1}^t  >
      m_\ell^t = m_{\ell-1}^{t+1}$, and $\ell \not\in \calB_{t+1}$.  Pick $\ell
      \in \calB_{t+1}$. We shall show this requires $\ell-1 \in \calB_t$.
      Since $\ell \not\in \calB_t$ as we just showed in (iii),
      \eqref{eq:mGeneral} yields:
      \begin{align}
	m_\ell^{t+1} & = \begin{cases}
	  m_{\ell}^t 	& \text{if $\ell+1 \not\in \calB_t$}, \\
	  m_{\ell+1}^t	& \text{if $\ell+1 \in \calB_t$},
	\end{cases} \label{eq:m1} \\ 
	m_{\ell-1}^{t+1} & = \begin{cases}
	  m_{\ell-2}^t	& \text{if $\ell-1 \in \calB_t$}, \\
	  m_{\ell-1}^t	& \text{if $\ell-1 \not\in \calB_t$}.
	\end{cases} \label{eq:m2}
      \end{align}
      Equation~\eqref{eq:m1}, together with \eqref{eq:bIndexOrder} imply
      $m_{\ell}^{t+1} \ge m_{\ell-1}^t$. Now if $\ell-1 \not\in \calB_t$, by
      \eqref{eq:m2}, $m_{\ell-1}^{t+1} = m_{\ell-1}^t \le m_{\ell}^{t+1}$. Hence
      $\ell \not\in \calB_{t+1}$.    
  \end{enumerate}

  It remains to show \eqref{eq:bIndexOrder} holds for all $t
  \ge 1$ by induction. The claim holds for $t=1$ by
  construction; $\calB_1 \subseteq \{2,4,\dots,2L-2\}$ (because of
  \eqref{eq:positiveIncrements}) and \eqref{eq:oldListSorted} is equivalent to
  \eqref{eq:bIndexOrder} for $t=1$.

  Pick $\ell \in \calB_{t+1}$. Assuming \eqref{eq:bIndexOrder} holds for $t$,
  we know $m_{\ell}^{t+1} \ge m_{\ell-1}^t$ (as we just showed). Furthermore,
  since $\ell-1 \in \calB_t$ (and $\ell-2 \not\in \calB_t$ due to
  \eqref{itm:nonAdjacentB}), \eqref{eq:mGeneral} yields $m_{\ell-2}^{t+1} =
  m_{\ell-1}^t \le m_{\ell}^{t+1}$.
\end{IEEEproof}

Property~\eqref{itm:forExec} means we can replace the condition of the
\textbf{if} block by $m_\ell^t \le m_{\ell-1}^t$ without changing the algorithm.
In other words, to determine whether we need to swap adjacent elements or not we
can take a look at the values stored at that positions at the beginning of each
round of the outer \textbf{while} loop. Furthermore,
property~\eqref{itm:nonAdjacentB} guarantees that each element, at each round,
participates in at most one swap operation. As a consequence the inner
\textbf{for} loop can be executed in parallel.  

Finally, property \eqref{itm:bPlusOne} together with the initial condition
$\calB_1 \subseteq \{2,4,\dots,2L-2\}$ implies that at odd rounds CAS
operations take place only between the even-indexed elements and their preceding
elements while at even rounds CAS operations take place only between the
odd-indexed elements and their preceding elements. 

\subsubsection{Implementation of Full Bubble Sorter}
\begin{figure}[t] 
  \centering 
  \includegraphics[width=0.42\textwidth]{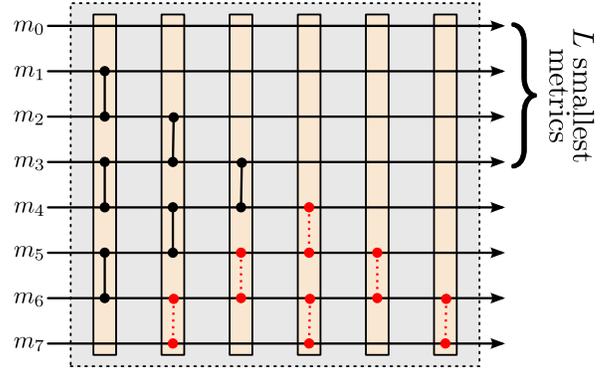} 
  \caption{Bubble sorter for $2L=8$. The full bubble sorter requires all the
  depicted CAS units, while in the simplified bubble sorter all CAS units in red
  dotted lines can be removed.}
  \label{fig:bubblesorter}
\end{figure}

Given the above considerations, we can implement the sorter in hardware as
follows: The sorter has $2L-2$ \emph{stages} each of them implementing a round
of bubble sort (i.e., an iteration of the \textbf{while} loop in
Alg.~\ref{alg:bubbleSort}).\footnote{In general the bubble sort terminates
  in up to $2L-1$ rounds but in our particular problem instance, since $m_0 =
\mu_0$ is the smallest element of the list, the first round is eliminated.}
At round $t$ of bubble sort the first $t$ elements are unchanged and the
sorter will have a \emph{triangular} structure.  Since in our setting, round $1$
is already eliminated, each stage $t$, $t = 1,2,\dots,2L-2$ only moves the
elements at indices $t, t+1, \dots, 2L-1$.  Each stage implements the execution
of the inner \textbf{for} loop in parallel using the required number of CAS
units. In Fig.~\ref{fig:bubblesorter} we show the structure of the sorter for
$2L=8$.  Using simple counting arguments we can show that the full bubble sort
requires $L(L-1)$ CAS blocks. 

\subsubsection{Implementation of Simplified Bubble Sorter}
So far we have only discussed about the implementation of a sorter that sorts
the entire list $\bm$. However, we only need the first $L$ ordered elements of
the list. Hence, we can simplify the full sorter as follows.
The first obvious simplification is to eliminate all the stages $L, L+1, \dots,
2L-2$ since we know that after round $L-1$ of the bubble sort, the first $L$
elements of the list correspond to an ordered list of the $L$ smallest elements
of the original list. Thus, the total number of required stages for this
simplified bubble sorter is
\begin{equation}
  s_{\rm tot}^{\rm B} = L-1. \label{eq:stagesb}
\end{equation}
Furthermore, we note that (due to property~\eqref{itm:nonAdjacentB}) each
element of the list at each round of the algorithm is moved at most by one
position. Consider the elements at positions $2L-t, 2L-t+1, \dots,
2L-1$ at round $t$. Since at most $L-t$ rounds of bubble sort are executed
(including the current round), these elements cannot be moved to the first half
of the list. Hence, we can eliminate the CAS units involving elements at indices
$2L-t, 2L-t+1 \dots, 2L-1$ at each stage $t = 1,2,\dots,L-1$ as well.
The simplified bubble sorter thus requires 
\begin{equation}
  c_{\rm tot}^{\rm B} = \frac12 L(L-1) \label{eq:compsb}
\end{equation}
CAS units. In Fig.~\ref{fig:bubblesorter} 
the parts of the sorter that can be eliminated are drawn with red dotted lines.

\section{Synthesis Results}
All results are obtained for the $90$~nm TSMC technology using the typical
timing library ($25^{\circ}$ C, $1.2$ V). The list elements are
assumed to be $Q = 8$ bits wide, as in \cite{Bala14tsp}. 

In Table~\ref{tab:r2l} we present synthesis results for the radix-$2L$ sorter of
\cite{Bala14} and the pruned radix-$2L$ sorter of \cite{Bala14tsp}. We observe
that the pruned radix-$2L$ sorter is at least $63\%$ smaller and at least $56\%$
faster than the full radix-$2L$ sorter.

\begin{table} 
  \centering 
  \begin{threeparttable}
    \caption{Synthesis Results for Radix-$2L$ and Pruned Radix-$2L$ Sorters}
    \label{tab:r2l} 
    \begin{tabular}{l|cc|cc} 
      & \multicolumn{2}{c}{Radix-$2L$} & \multicolumn{2}{c}{Pruned Radix-$2L$}
      \\ 
      \hline 
      & Freq. (MHz)	& Area ($\mu\text{m}^2$)	
      & Freq. (MHz)	& Area ($\mu\text{m}^2$) \\ 
      \hline 
      $L=2$	& $2128$ 	& $3007$ 	& $4545$	& $608$ \\ 
      $L=4$	& $1111$	& $12659$	& $2083$	& $3703$ \\ 
      $L=8$	& $526$		& $50433$	& $1031$	& $18370$ \\ 
      $L=16$	& $229$		& $238907$	& $372$		& $70746$ \\ 
      $L=32$	& n/a\tnote{*}	& n/a\tnote{*}	& $145$		& $376945$ \\ 
    \end{tabular}
    \begin{tablenotes}
    \item[*] For $L=32$ the synthesis tool ran out of memory on a machine with
      48GB of RAM, most likely due to the lack of structure in the circuits that
      generate the control signals for the multiplexers in the radix-$2L$
      sorter.
    \end{tablenotes}
  \end{threeparttable}
\end{table}

In Table~\ref{tab:bitonic} we present synthesis results for the bitonic sorter
of \cite{Lin14} and the pruned bitonic sorter presented in this paper. 
We observe that, as discussed in Section~\ref{sec:pbs}, the improvement in terms
of both area and operating frequency are diminishing as the list size $L$ is
increased.  Nevertheless, even for $L=32$ the pruned bitonic sorter is $5\%$
faster and $14\%$ smaller than the full bitonic sorter.
\begin{table} 
  \caption{Synthesis Results for Bitonic and Pruned Bitonic Sorters}
  \label{tab:bitonic} 
  \centering 
  \begin{tabular}{l|cc|cc} 
    & \multicolumn{2}{c}{Bitonic} & \multicolumn{2}{c}{Pruned Bitonic}  \\ 
    \hline 
    & Freq. (MHz)		& Area ($\mu\text{m}^2$)	
    & Freq. (MHz)		& Area ($\mu\text{m}^2$) \\ 
    \hline 
    $L=2$	& $1370$	& $2109$	& $4545$	& $608$  \\ 
    $L=4$	& $676$		& $8745$	& $952$		& $3965$  \\ 
    $L=8$	& $347$		& $27159$	& $478$		& $20748$ \\ 
    $L=16$	& $214$		& $82258$	& $256$		& $69769$ \\ 
    $L=32$	& $157$		& $238721$	& $166$		& $205478$ \\ 
  \end{tabular}
\end{table}

In Table~\ref{tab:allsorters} we present synthesis results for the simplified
bubble sorter described in Section~\ref{sec:bubble_sorting}.  We observe that,
for $L \leq 8$, the simplified bubble sorter has a lower delay than the pruned
bitonic sorter. This happens because, as can be verified by evaluating
\eqref{eq:stagesbt} and \eqref{eq:stagesb}, for $L \leq 8$ the bubble sorter has
fewer stages than the pruned bitonic sorter while for $L > 8$ the situation is
reversed.  Similar behavior can be observed for the area of the sorters, where
the bubble sorter remains smaller than the pruned bitonic sorter for $L \leq
16$.

We also observe that, for $L \leq 16$, the pruned radix-$2L$ sorter is faster
than the other two sorters and similar in area to the pruned bitonic sorter,
while the simplified bubble sorter is significantly smaller.  Thus, for $L \leq
16$ the pruned bitonic sorter is not a viable option, while trade-offs between
speed and area can be made by using either the pruned radix-$2L$ sorter or the
simplified bubble sorter.  For $L=32$, however, the pruned bitonic sorter has a
higher operating frequency \emph{and} a smaller area than the other two sorters.

\begin{table}[t] 
  \centering
  \begin{threeparttable} 
    \caption{Comparison of Pruned Radix-$2L$, Pruned Bitonic, and Simplified
    Bubble Sorters}
    \label{tab:allsorters} 
    \begin{tabular}{l|cc|cc|cc} 
      & \multicolumn{2}{c}{Pruned Radix-$2L$} 	
      & \multicolumn{2}{c}{Pruned Bitonic}	
      & \multicolumn{2}{c}{Simplified Bubble} \\ 
      \hline 
      & Freq.	& Area			& Freq.	& Area		
      & Freq.	& Area \\ 
      & (MHz)	& ($\mu\text{m}^2$)	& (MHz)	& ($\mu\text{m}^2$)		
      & (MHz)	& ($\mu\text{m}^2$) \\ 
      \hline 
      $L = 2$\tnote{*}
      & $4545$		& $608$		& $4545$	& $608$ 
      & $4545$		& $608$ \\ 
      $L = 4$		
      & $2083$		& $3703$	& $952$		& $3965$
      & $1388$ 		& $2756$ \\ 
      $L = 8$		
      & $1031$		& $18370$	& $478$		& $20748$ 	
      & $534$		& $11726$ \\ 
      $L = 16$	
      & $372$		& $70746$	& $256$		& $69769$
      & $247$ 		& $51159$ \\ 
      $L = 32$	
      & $145$		& $376945$	& $166$		& $205478$ 
      &	$127$		& $212477$ 
    \end{tabular}
    \begin{tablenotes}
    \item[*] For $L=2$ it can easily be seen that all three sorters are
      equivalent.
    \end{tablenotes}
  \end{threeparttable}
\end{table}

\section{Conclusion}
In this work, we presented a pruned bitonic and a bubble sorter that exploit the
structure of the elements that need to be sorted in SCL decoding of polar codes.
Our results indicate that the bitonic sorter used in \cite{Lin14}, even with the
pruning proposed in this work, is not a suitable choice for list sizes $L \leq
16$. Our simplified bubble sorter, on the other hand, provides a meaningful
trade-off between speed and area with respect to the pruned radix-$2L$ sorter
used in~\cite{Bala14tsp} (which still remains the fastest sorter for $L \leq
16$).  For $L=32$, however, the superior delay and area scaling of the pruned
bitonic sorter make it $14\%$ faster and $45\%$ smaller than the second-fastest
radix-$2L$ sorter. Moreover, both the pruned bitonic and the bubble sorter have
stages that are identical in terms of delay, thus enabling much simpler
pipelining than the radix-$2L$ sorter.

\section*{Acknowledgment}
This work was supported by the Swiss NSF under grant numbers 200021\_149447 and
200020\_146832. 
 


\bibliographystyle{IEEEtran}
\bibliography{IEEEabrv,refs}
%

\end{document}

%% file: notation.tex
\newcommand{\bm}{\mathbf{m}}

\newcommand{\calA}{\mathcal{A}}
\newcommand{\calB}{\mathcal{B}}


%% file: definitions.tex
\newtheorem{theorem}{Theorem}

\newtheorem{lemma}[theorem]{Lemma}

\newtheorem*{theorem*}{Theorem}
\newtheorem*{exercise*}{Exercise}
\newtheorem*{corollary*}{Corollary}
\newtheorem*{lemma*}{Lemma}
\newtheorem*{property*}{Property}
\newtheorem*{proposition*}{Proposition}
\newtheorem*{problem*}{Problem}
\newtheorem*{observation*}{Observation}

\theoremstyle{definition}

\theoremstyle{remark}
\newtheorem*{remark}{Remark}